\documentclass{aastex631}
\usepackage{graphicx}

\begin{document}

\title{The Effect of Work Function on Dust Charging and Dynamics on the Airless Celestial Body}

\author[0000-0002-4321-9008]{Ronghui Quan}
\affiliation{Nanjing University of Aeronautics and Astronautics, Nanjing, Jiangsu, 210016, People's Republic of China}

\author[0000-0002-9859-0277]{Zhigui Liu}
\affiliation{Nanjing University of Aeronautics and Astronautics, Nanjing, Jiangsu, 210016, People's Republic of China}

\author[0000-0003-3498-2172]{Zhiying Song}
\affiliation{Nanjing University of Aeronautics and Astronautics, Nanjing, Jiangsu, 210016, People's Republic of China}

\correspondingauthor{Zhigui Liu}
\email{lzgyyy@nuaa.edu.cn}


\begin{abstract}
The charged dust on the surface of airless celestial bodies, such as the moon and asteroids, is a threat to space missions. Further research on the charged dust will contribute to the success of space missions. In this paper, we study the charging and dynamics of dust particles with different work functions. By integrating the photoelectron energy distribution function over four illuminated areas with different work functions, we evaluated the photoelectron concentration in these four areas. At each area, using the photoelectron concentration, we solve the dust charging and dynamics equations with two different gravitational acceleration values. The results reveal that the dust with a larger work function can reach higher equilibrium states. These states include dominant photoelectron-related charging currents, charge numbers, and levitation heights. We suggest that the equilibrium states all hold a clear inverse relationship with the work functions of dust particles when the solar zenith angle varies from $0^{\circ}$ to $90^{\circ}$, displaying consistent trends under different gravitational accelerations. We also find that dust particles seem unable to stably levitate at a critical solar zenith angle. The value of this critical SZA follows the same rule subjected to the work function.
\end{abstract}




\section{Introduction} \label{sec:intro}

The charged dust on the surface of airless bodies is a serious threat to human exploration of these bodies. On airless bodies like the Moon or asteroids, dust becomes charged due to direct interactions with solar wind and solar radiation, and can be levitated due to electrostatic force. This could lead to visual impairment, dust adhesion and contamination, mechanical blockages, surface erosion, and physiological irritation\citep{2022ChPhB..31d5201X}. To avoid these risks, theories on dust particle charging and dynamics have been developed.

A typical example of charged dust on airless celestial bodies is lunar dust. Since the observation of the horizon glow on the moon\citep{1973ASSL...37..545C}, many studies on the dust charging mechanism of lunar dust have been published. Some earlier researches suggest that the potential and electric field within the sheath may support the dynamics and levitation of the charged particles\citep{1971JGR....76.2498G,1992EM&P...56....7N,1998JGR...103.6605N}. Recently, “the patched charge model” of dust particle charging established through experiments, appears to be a breakthrough\citep{2016GeoRL..43.6103W,2017GeoRL..44.3059S,2020PhPl...27e2901M}. The theoretical model based on the “the patched charge model” has also been established\citep{2016JGRE..121.2150Z}, and dust dynamics on asteroids have been studied by using this model\citep{2020MNRAS.496L..80M,2022PSJ.....3...85H}.

Due to the similarity between asteroids and the Moon, theories on the charging and dynamics of lunar dust are also applicable to asteroids. The lunar charging theory has been used to explain the formation of dust ponds on Eros\citep{colwell2005dust,2008Icar..195..630H}. Furthermore, surface charging processes may also occur on comets\citep{2015P&SS..119...24N}. With a smaller size, the cost of simulating interactions between asteroids and solar wind plasma is less. Consequently, many software or codes have been developed to study the charging of asteroids and the dust dynamics on their surfaces. For instance, particle-in-cell simulations have been conducted on irregularly-shaped asteroids or dust in the solar wind\citep{zhao2022kinetic,2023NatSR..13.1111D,lund2023charging}, and using the Spacecraft Plasma Interaction Software(SPIS) for smooth and regular asteroids like 2016H03\citep{2023ApJ...952...61X}.

Both past and present studies indicate that the ability of airless celestial body surface or dust particles to emit photoelectrons depends on the energy of the incident photons and the material's work function\citep{2014JETPL..99..115P,2020PhPl...27h2906M}. However, many studies on the charging of dust particles on the lunar or asteroids are based on the assumption of dust particles with a single work function of 5eV or 6eV for simplicity. Yet, Experimental studies on lunar regolith brought back by the Apollo missions suggest that the work function of lunar regolith is around 5eV-6eV\citep{1972LPSC....3.2655F,2002JGRE..107.5105S,colwell2007lunar}. And according to existing astronomical exploration results, carbonaceous and silicate asteroids dominate, exhibiting diverse materials including plagioclase, pyroxene, and ilmenite\citep{2022ApJ...935L...3B}. This implies that considering more material types in calculating the charging currents related to photoelectrons on the surface of airless celestial bodies will lead to different charging and dynamics results.

In this paper, we recalculated the surface charging currents related to the work functions of four types of dust particles and obtained their subsequent charging and dynamics results. It is worth noting that we used a novel photoelectron yield. Our findings revealed a dependence of dust particle charging and dynamics results on the work function, contributing to a better understanding of the dust environment on the surface of airless celestial bodies.

The structure of this paper is as follows: Section \ref{sec:model} introduces the relevant theories, including the new model for predicting photoelectron yields, methods for calculating photoelectron concentrations, and charging and dynamics equations. Section \ref{sec:results} is the numerical results and discussion. Section \ref{sec:conclusion} presents our conclusion.

\section{MODEL DESCRIPTION} \label{sec:model}

\subsection{Photoelectron Yield of Dust Particles}
When sunlight irradiates dust particles, these particles emit and absorb photoelectrons, leading to two critical currents influencing the charging and dynamics: the photoelectron collection current and the photoemission current. Estimating these two currents necessitates the photoelectron yield defined as the number of photoelectrons ejected from a surface per absorbed photon. Photoelectric emission experiments\citep{1972LPSC....3.2655F} conducted on lunar soil samples brought back by the Apollo mission revealed that the samples exhibit a peak photoelectron yield of approximately 0.09 at an electromagnetic wavelength of around 90nm, corresponding to a photon energy of 13.8eV. Additionally, the yield decreases to about $10^{-6}$ when the photon energy approaches the work function of the samples, which is estimated to be 5eV-6eV. However, the effective wavelength range of the experiment is narrow, spanning only 50nm to 250nm, which limits its applicability. Additionally, it is difficult to avoid the influence of the atmosphere of the Erath on the lunar sample\citep{2020PhPl...27h2906M}. Consequently, some studies prefer to use the semi-empirical photoelectron yield for lunar surface material\citep{1973JGR....78.3668W}:
\begin{equation}
\label{eq.1}
Y_{\text {semi }}=\left\{\begin{array}{l}
0 \quad\left(E_{ph}<6 e V\right) \\
0.02+0.06\left(E_{ph}-6\right) \quad\left(6eV \leq E_{ph} \leq 9eV\right) \\
0.2 \quad\left(E_{ph}>9 e V\right) .
\end{array}\right.
\end{equation}

Although the semi-empirical method offers valuable insight into photoelectron yield across a broad wavelength range, it cannot be used to predict the photoelectron yield of materials with other work functions. To address this issue, we use a totally theoretical model particularly developed for astronomical silicates in this paper. Compared to the semi-empirical ones, this method  is more reliable in astronomical research\citep{2016MNRAS.459.2751K}:

\begin{equation}
\label{eq.2}
Y=\frac{1}{2}\left[1-\sqrt{\frac{W}{E_{p h}}}+\frac{l_{a}}{l_{e}} \log \left(\frac{\sqrt{W/E_{ph}}+l_{a}/l_{e}}{1+l_{a}/l_{e}}\right)\right],
\end{equation}
where $W$ is the work function of dust particles, $l_a$ is the photon attenuation length, and $l_e$ is the mean free path of electron, both  $l_a$ and $l_e$ are dependent on the photon energy $E_{ph}$\citep{seah1979quantitative}.

\subsection{Photoelectron Concentration} 
Photoelectrons are produced near the illuminated surface of airless celestial bodies through interactions with solar radiation. The concentration of photoelectrons can be determined by integrating their distribution function. Early studies on dust charging on the lunar surface assumed a Maxwellian distribution of photoelectrons. Recent research has indicated that the distribution of photoelectrons within the photoelectron sheaths does not strictly adhere to the Maxwell distribution\citep{2015P&SS..116...18S,2020PhPl...27h2906M}. In this paper, Popel et al's method was used to recalibrate the distribution function and obtain the concentration of photoelectrons emitted from dust particles and surfaces of the airless celestial body\citep{2014JETPL..99..115P}. The energy distribution function of photoelectrons in an illuminated area with work function $W$ is as follows:
\begin{equation}
\label{eq.3}
f\left(E_{e}\right) d E_{e}=2 \cos (\theta) \sqrt{\frac{2 m_{e}}{E_{e}}} \int_{E_{e}+W}^{\infty} Y\left(E_{p h}\right) S\left(E_{p h}\right) d \rho d E_{p h},
\end{equation}
where $\theta$ is the solar zenith angle, $E_e$ is the photoelectron energy, $m_e$ is the electron mass, $S(E_{ph})$is the solar radiation spectrum, and $d\rho$ represents the probability of an electron with energy $E_e$ being emitted within the energy range of $dE_e$ due to the absorption of a photon with the energy of $E_{ph}$,
\begin{equation}
d\rho=\frac{6\left(E_{ph}-W-E_{e}\right)}{\left(E_{ph}-W\right)^{3}} E_{e}dE_{e}\quad0\leq E_{e} \leq E_{ph}-W,
\end{equation}

The concentration $N_0$ in an illuminated area which the particles are located can be obtained by integrating the distribution function, 
\begin{equation}
\label{eq.5}
N_{0}=\int_{0}^{\infty} f\left(E_{e}\right) d E_{e}.
\end{equation}

\begin{figure}[ht!]
\plotone{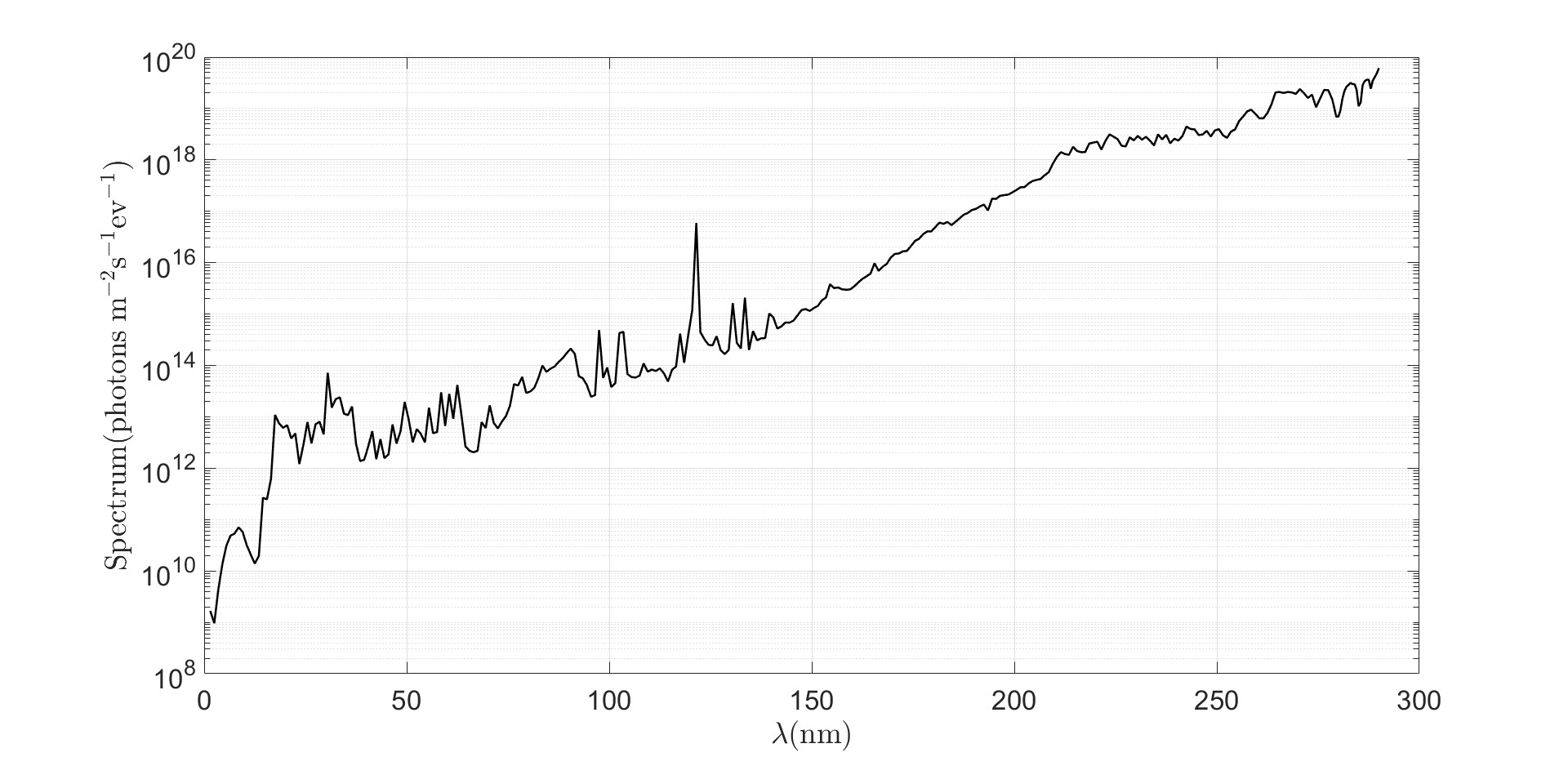}
\caption{Wavelength dependence of the solar radiation spectrum under average solar activity conditions.\label{fig:spectrum}}
\end{figure}

To avoid the influence of solar activities on the calculation results, such as solar flares. We use solar radiation spectrum under average solar activity conditions\citep{2004SoEn...76..423G}, as shown in Figure \ref{fig:spectrum}.

\subsection{The Charging and Dynamics Equations}
On the day side of an airless celestial body, such as an asteroid or the Moon, both dust particles and the celestial body's surface generally acquire positive charges primarily owing to the dominant photoemission current. Consequently, electrons are attracted to the positively charged surface, leading to the formation of an electric field $E$ nearly perpendicular to the surface of the celestial body, directed in an anti-centripetal direction\citep{2020PlPhR..46..527Z}:
\begin{equation}
E(h,\theta)=\frac{2kT_{eph}}{e} \frac{\sqrt{0.5\cos\theta}}{\lambda_{D}+h \sqrt{0.5\cos\theta}},
\end{equation}
where $\lambda_D=\sqrt{\varepsilon_0kT_{eph}/(e^2N_0)}$ is Debye length at the surface.

As a result, the positively charged dust particles will be subjected to the electrostatic force and may overcome the gravity and cohesive force binding them to the ground, leading to levitation. During levitation, dust particles continue to accumulate charge until a charging equilibrium is reached. This paper only focuses on the height variation of dust particles subjected to gravity and electrostatic force:
\begin{equation}
\label{eq.7}
m_{d} \frac{d^{2} h}{d t^{2}}=Q E-m_{d} g_{a},
\end{equation}
\begin{equation}
\label{eq.8}
\frac{d Q}{d t}=I_{p h}+I_{i}-I_{p h e}-I_{e}.
\end{equation}
By setting both Eq.(\ref{eq.7}) and Eq.(\ref{eq.8}) to zero, we can obtain the equilibrium equations:
\begin{equation}
\label{eq.9}
h_{\text{equibrium }}=\frac{2 k T_{eph} Q_{\text{equibrium }}}{m_{d} g_{a}e}-\frac{\lambda_{D}}{\sqrt{0.5\cos(\theta)}},
\end{equation}
\begin{equation}
0=I_{p h}+I_{i}-I_{p h e}-I_{e}.
\end{equation}
where $m_d$ is the mass of the dust particles, $h$ is the vertical distance of the dust particle from the surface of the airless celestial body, $E$ is the electric field on the surface, and $g_a$ is the gravitational acceleration.

Four main currents that flow through the surface of a dust particle are considered: Photoemission current $I_{ph}$, Photoelectron collection current $I_{phe}$, solar wind electron current $I_e$, and solar wind ion current $I_i$, they take the following form\citep{2020PlPhR..46..527Z}:
\begin{equation}
\label{eq.11}
I_{ph}=\pi r^{2}e N_{0}\sqrt{\frac{kT_{eph}}{2\pi m_{e}}}\left(1+\frac{Qe}{CkT_{eph}}\right)\exp\left(-\frac{Qe}{CkT_{eph}}\right),
\end{equation}
\begin{equation}
\label{eq.12}
I_{phe}=\pi r^{2} e n_{pe} \sqrt{\frac{8 k T_{eph}}{\pi m_{e}}}\left(1+\frac{Q e}{C k T_{e p h}}\right),
\end{equation}
\begin{equation}
I_{e}=\pi r^{2}e n_{es} \sqrt{\frac{8kT_{es}}{\pi m_{e}}}\left(1+\frac{Q e}{CkT_{es}}\right),
\end{equation}
\begin{equation}
I_{i}= \pi r^{2} e n_{i s} \sqrt{\frac{8 k T_{i s}}{\pi m_{i}}} \exp \left(-\frac{Q e}{C k T_{i s}}\right).
\end{equation}
where $r$ is the radius of a dust particle, $k$ is the Boltzmann constant, $T_{eph}$ is the temperature of photoelectron, which typically takes the value of $kT_{eph}=2.2eV$\citep{colwell2007lunar}, $Q$ is the charge of a dust particle, e is the elementary charge, $C=4 \pi \varepsilon_0 r$ is the capacitor of a spherical dust particle, $\varepsilon_0$ is the permittivity of free space, $m_e$ and $m_i$ are the mass of electron and ion, respectively. $n_{pe}$ represents the photoelectron concentration, which inversely varies with the height $h$ perpendicular to the surface of the airless celestial body. The temperature and concentration of electrons and ions in the solar wind are dependent on the heliocentric distance d(AU) of the celestial body from the sun\citep{1996Icar..124..181L}:
\begin{equation}
n_{e s}=n_{i s}=5 \times 10^{6} d^{-2},
\end{equation}  
\begin{equation}
T_{e s}=1.5 \times 10^{5} d^{-2 / 3},
\end{equation}  
\begin{equation}
T_{i s}= 10^{5} d^{-2 / 3},
\end{equation}  
where $n_{es}$ and $n_{is}$ both are in $m^{-3}$, $T_{es}$ and $T_{is}$ both are Kelvin temperature.

To facilitate the levitation of dust particles, the distance between the airless celestial body and the Sun must be carefully chosen. If it is too close, the dust particles will escape due to the Sun's gravitational pull, on the other hand, if it is too far away from the Sun, the electrostatic force associated with the concentration of photoelectrons may be too small to allow the levitation of dust particles. Studies have shown that dust particles are more likely to be levitated when the airless celestial body is located within 1AU to 2AU from the Sun\citep{colwell2005dust,2020P&SS..18004775L,2022PSJ.....3...85H}. Hence, we take the aphelion distance 1.78AU of Eros as the value of d.

\section{RESULTS and discussion} \label{sec:results}
\begin{figure}[htbp]
    \centering
    \begin{minipage}{0.43\textwidth}
        \centering
        \includegraphics[width=\textwidth]{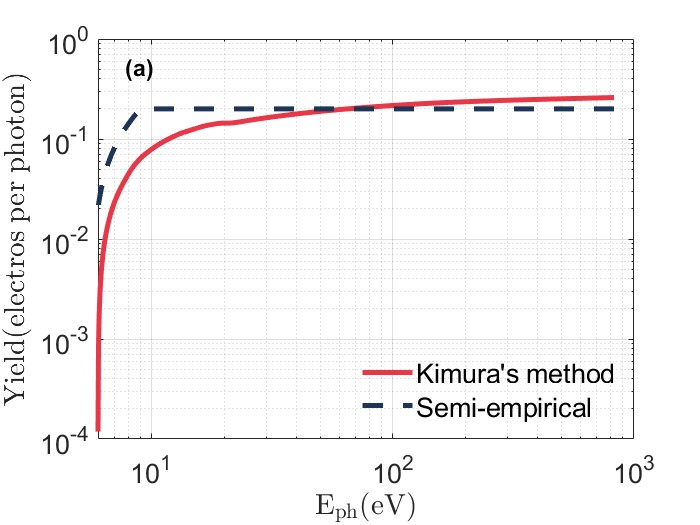}
    \end{minipage}
    \begin{minipage}{0.43\textwidth}
        \centering
        \includegraphics[width=\textwidth]{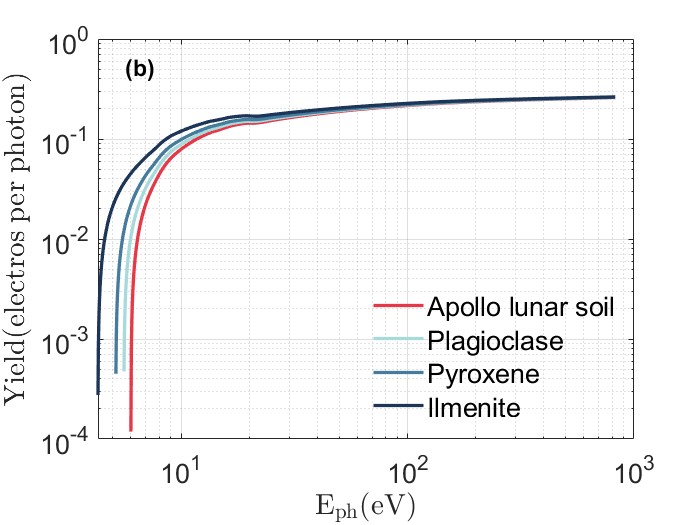}
    \end{minipage}
   \caption{(a)Photoelectron yield for Apollo lunar soil. The black dashed line represents the semi-empirical yield(Eq.(\ref{eq.1})), and the red solid line represents the yield of Kimura’s method(Eq.(\ref{eq.2})). (b)Photon energy dependence of photoelectron yield of four different types of dust particles using Kimura’s method. Red line represents Apollo lunar soil, cyan line represents plagioclase, blue line represents pyroxene, and deep blue line represents ilmenite.}
   \label{fig.2}
\end{figure}

Since Eqs.(\ref{eq.3})-(\ref{eq.5}) are methods developed for dust particles with a single work function to estimate the photoelectron concentration in the area which they are located in, we assume that there are four areas with the same illuminated conditions on the airless celestial body, disregarding any effects due to rotation and topographical variation. The dust particles in each aera consist of Apollo lunar soil, plagioclase, pyroxene, and ilmenite, respectively. The work functions of these four types of dust particles are shown in Table \ref{tab.1}\citep{Nature...41572037}. We first present the photoelectron yield for Apollo lunar soil using the semi-empirical method(Eq.(\ref{eq.1})) and the theoretical ones(Eq.(\ref{eq.2})), respectively, as shown in Figure \ref{fig.2}(a). The results demonstrate a reasonable agreement between the semi-empirical method and the theoretical ones, indicating the rationality of using the theoretical method strategy. The theoretical photoelectron yield for four different types of dust particles is illustrated in Figure \ref{fig.2}(b).

Let $\theta=0^{\circ}$, which corresponds to the noon time on the airless celestial body, the photoelectron concentration $N_{e0}$ of four areas at the noon time can be obtained by integrating Eqs.(\ref{eq.3})-(\ref{eq.5}), and notice that $N_0=N_{e0}\cos(\theta)$. The results are shown in Table \ref{tab.1}.
\begin{deluxetable}{ccc}[h]
\tablecaption{Photoelectron concentration of four aeras at noon \label{tab.1}}
\tablehead{
\colhead{Material} & \colhead{Work Function} & \colhead{Concentration}\\
\colhead{} & \colhead{(eV)} & \colhead{($m^{-3}$)}} 
\startdata 
{Apollo Lunar Soil}  &  6.00 & 2.4501E10 \\ 
{Plagioclase}        &  5.58 & 9.7963E10 \\ 
{Pyroxene}           &  5.14 & 3.1891E11 \\
{Ilmenite}           &  4.29 & 2.3962E12 \\
\enddata
\end{deluxetable}

The result of Popel for the photoelectron concentration of the Apollo lunar soil is $2\times 10^{11} m^{-3}$, and ours is an order of magnitude lower\citep{2013SoSyR..47..419P}, This discrepancy can be attributed to the fact that when the photon energy is less than 100eV, the semi-empirical formula has a higher yield compared to the formula we used, particularly near the work function, the former is almost two orders of magnitude larger than the latter\citep{2020PhPl...27h2906M,2020PlPhR..46..527Z}, resulting in the acceptable difference. Table \ref{tab.1} shows that the concentration in the four regions increases with the decrease in the work function.

To neglect the effects of the mass of dust particles, we assume that four types of dust particles weigh equally. Considering the possible effects of gravitational accelerations on dust particles due to the unevenness of gravity on the airless celestial body or the fact that dust particles may be on different celestial bodies, we examined the subsequent charging and dynamics by using gravitational acceleration values of two classic airless celestial bodies---Vesta and the Moon, which are $0.25m\;s^{-2}$ and $1.63m\;s^{-2}$\citep{2012Sci...336..684R,2014Icar..240..103K}, respectively. The reason for this choice is that these two bodies have relatively high surface gravitational accelerations so that the particles are unlikely to escape the gravitational pull during their motion when we are primarily interested in the dynamics near the surface of the airless celestial body.

\begin{figure}[ht]
    \centering
    \begin{minipage}{0.43\textwidth}
        \centering
        \includegraphics[width=\textwidth]{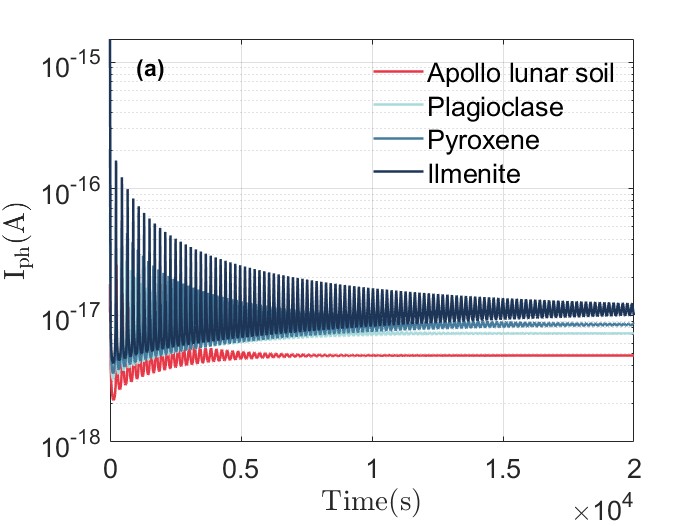}
    \end{minipage}
    \begin{minipage}{0.43\textwidth}
        \centering
        \includegraphics[width=\textwidth]{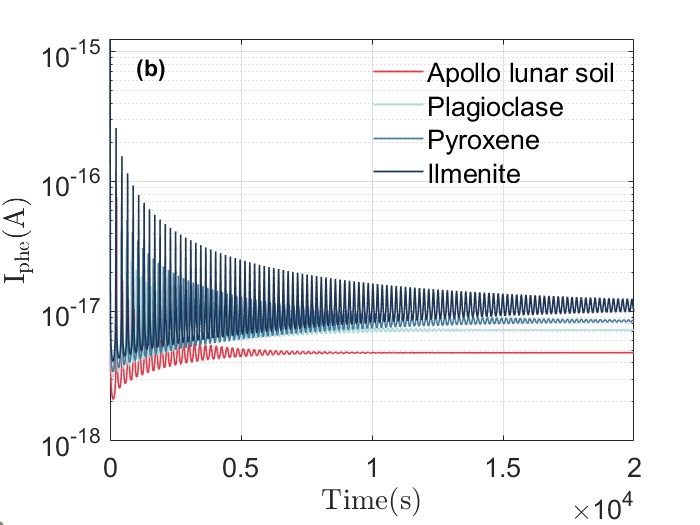}
    \end{minipage}
    \begin{minipage}{0.43\textwidth}
        \centering
        \includegraphics[width=\textwidth]{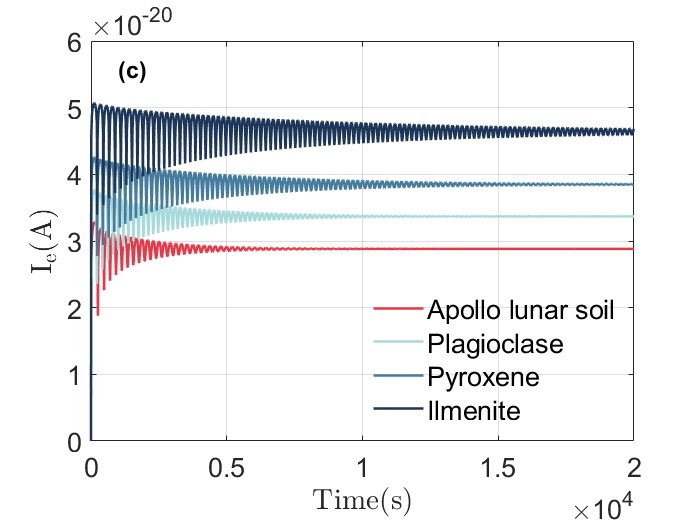}
    \end{minipage}
    \begin{minipage}{0.43\textwidth}
        \centering
        \includegraphics[width=\textwidth]{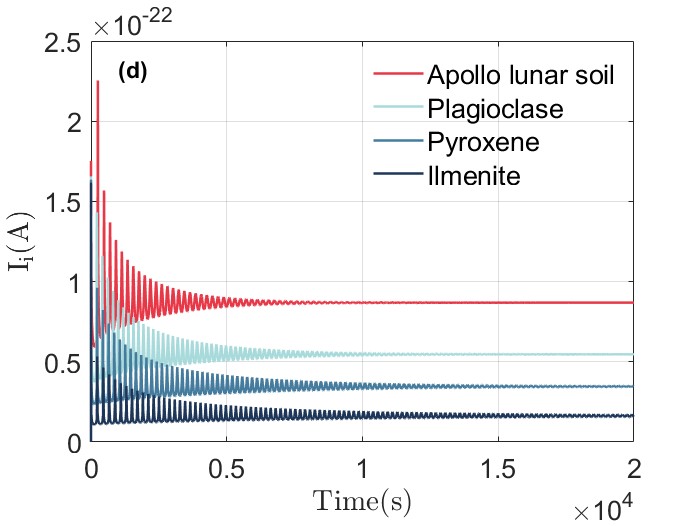}
    \end{minipage}
   \caption{Charging currents of lofted dust particles under $g_a=0.25m\;s^{-2}$, $\theta=0^{\circ}$. (a) Photoemission current, (b) Photoelectron collection current, (c) Solar wind electron current, (d) Solar wind ion current.}
   \label{fig.3}
\end{figure}

\begin{figure}[t]
    \centering
    \begin{minipage}{0.43\textwidth}
        \centering
        \includegraphics[width=\textwidth]{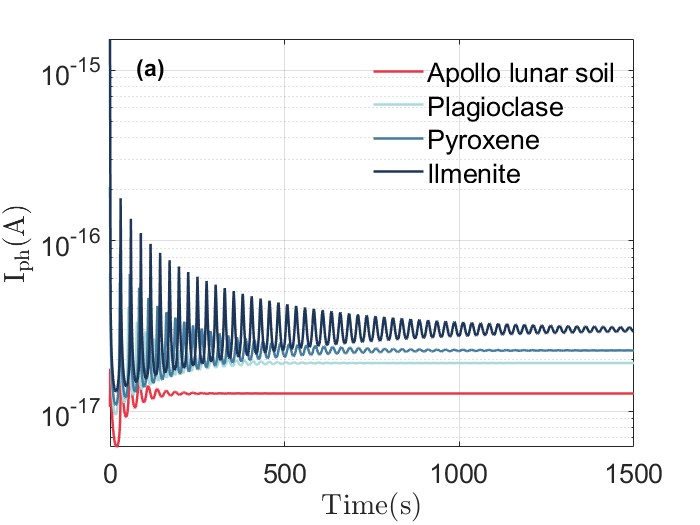}
    \end{minipage}
    \begin{minipage}{0.43\textwidth}
        \centering
        \includegraphics[width=\textwidth]{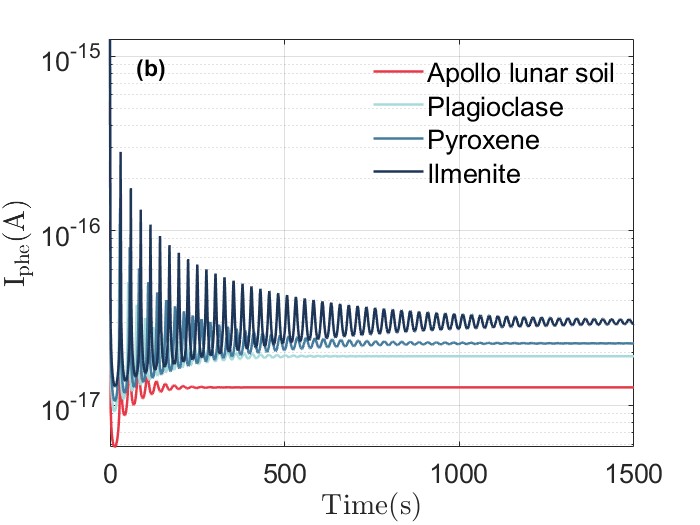}
    \end{minipage}
    \begin{minipage}{0.43\textwidth}
        \centering
        \includegraphics[width=\textwidth]{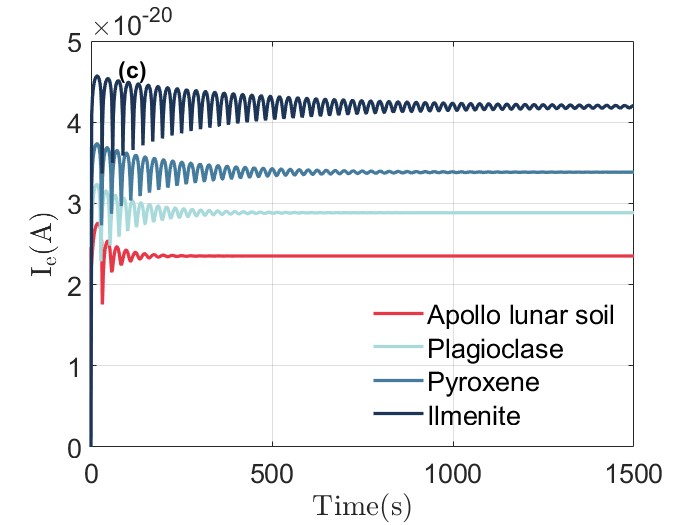}
    \end{minipage}
    \begin{minipage}{0.43\textwidth}
        \centering
        \includegraphics[width=\textwidth]{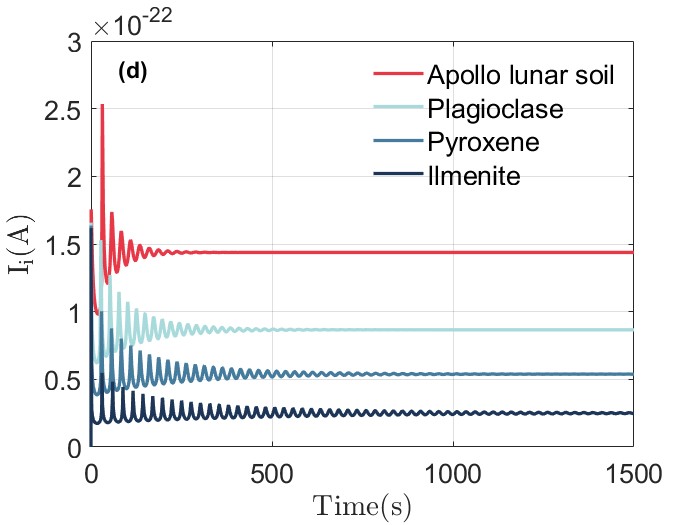}
    \end{minipage}
   \caption{Charging currents of lofted dust particles under $g_a=1.63m\;s^{-2}$, $\theta=0^{\circ}$. (a) Photoemission current, (b) Photoelectron collection current, (c) Solar wind electron current, (d) Solar wind ion current.}
   \label{fig.4}
\end{figure}

According to Eq.(\ref{eq.11}) and Eq.(\ref{eq.12}), the difference in the concentration as shown in Table \ref{tab.1} would directly affect the charging results of the dust particles in the four areas. Figures \ref{fig.3}-\ref{fig.4} illustrate the charging currents of lofted dust particles at noon with $0.25m\;s^{-2}$ and $1.63m\;s^{-2}$, respectively. Figures \ref{fig.3} and \ref{fig.4} both show that all currents reach equilibrium after a period of oscillation, and dust particles with a larger work function exhibit smaller equilibrium for photoelectron-related and solar wind electron currents than those with lower ones, while higher solar wind ion currents. The disparity in photoemission currents that is shown in Figure \ref{fig.3}(a) and Figure \ref{fig.4}(a) can be explained by the proportional relationship with the concentration of photoelectrons, as indicated by Eq.(\ref{eq.11}), which suggests that dust particles with lower work functions exhibit a higher equilibrium photoemission current.

Owing to the dust particles with lower work function can emit more photoelectrons to obtain more positive charges as shown in Table \ref{tab.1}, they can attract more photoelectrons and solar wind electrons, resulting in higher equilibrium photoelectron collection currents and solar wind electron currents, as illustrated in Figures \ref{fig.3}(b)-(c) and Figures \ref{fig.4}(b)-(c). Simultaneously, they would repel more solar wind ions, leading to a smaller equilibrium solar wind ion current compared to dust particles with higher work function, as illustrated in Figure \ref{fig.3}(d) and Figure \ref{fig.4}(d).

\begin{figure}[ht]
    \centering
    \begin{minipage}{0.90\textwidth}
        \centering
        \includegraphics[width=\textwidth]{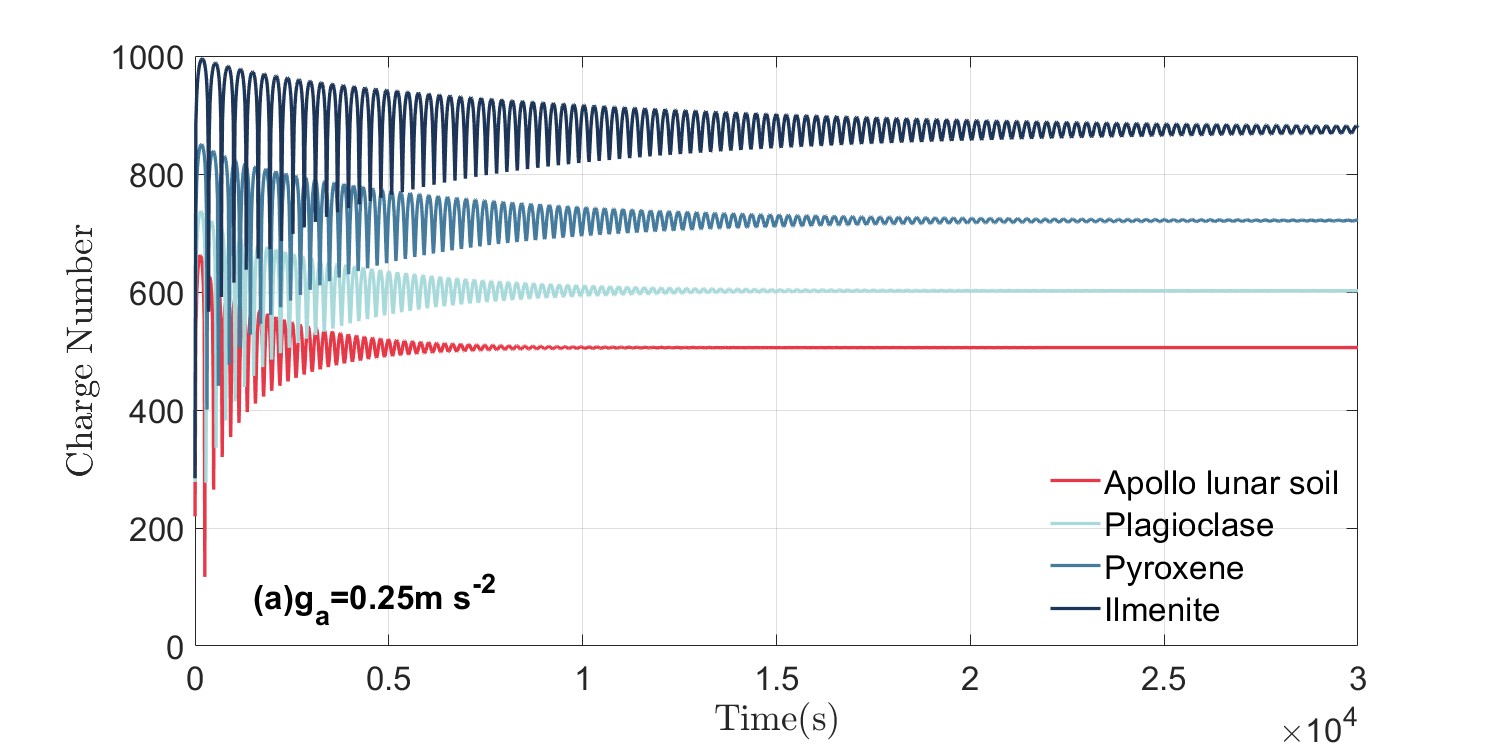}
    \end{minipage}
    \begin{minipage}{0.9\textwidth}
        \centering
        \includegraphics[width=\textwidth]{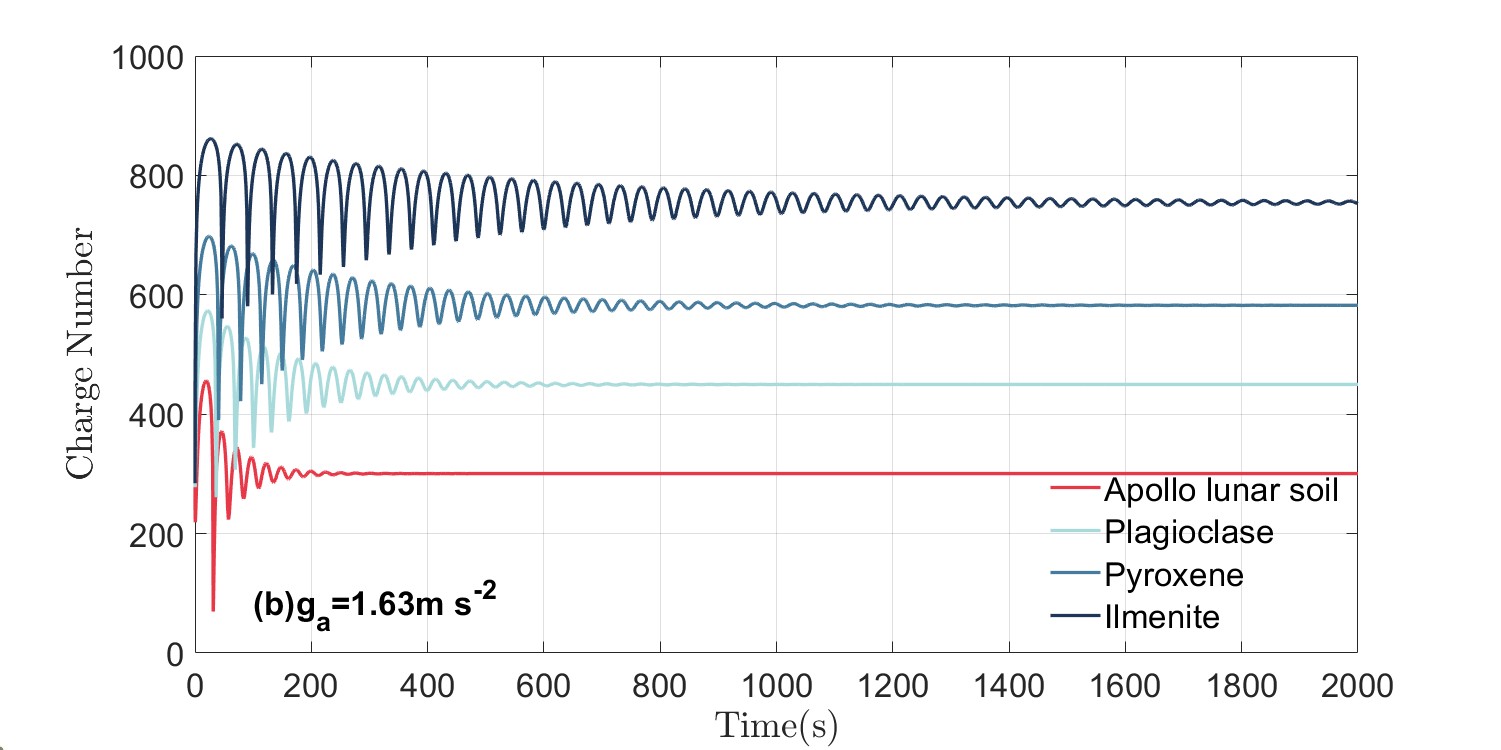}
    \end{minipage}
   \caption{(a) The charge numbers of lofted dust particles as a function of time with $g_a=0.25m\;s^{-2}$, $\theta=0^{\circ}$. (b) The charge numbers of lofted dust particles as a function of time with $g_a=1.63m\;s^{-2}$, $\theta=0^{\circ}$.}
   \label{fig.5}
\end{figure}

Figure \ref{fig.5} illustrates the charge numbers $Z(Q=Ze)$ of lofted dust particles at noon with $g_a=0.25m\;s^{-2}$ and $g_a=1.63m\;s^{-2}$, respectively. Figure \ref{fig.5} shows that the charge numbers also oscillate and eventually stabilize. The surface equilibrium charges of the four types of dust particles can reach several hundred elementary charges with two gravity values, which is similar to the results of Popel\citep{2022PhPl...29a3701P}. With a high gravity($g_a=1.63m\;s^{-2}$), the equilibrium charge numbers are as follows: 750 for Ilmenite, 580 for pyroxene, 450 for plagioclase, and 300 for Apollo lunar soil, whereas with a low gravity($g_a=0.25m\;s^{-2}$), the values are approximately 875, 720, 600, and 500. Due to the charge numbers being primarily associated with the dominant photoemission currents, the charge numbers should follow the same relationship with the work function as the photoemission currents do. As expected, the results indicate that equilibrium charge numbers decrease with increasing work functions. As mentioned above, this is due to dust particles with a lower work function can emit more photoelectrons as shown in Table \ref{tab.1}, which allows a higher photoemission current flowing through them. Hence, they can accumulate more positive charges at the same time.
\clearpage

\begin{figure}[ht]
    \centering
    \begin{minipage}{0.90\textwidth}
        \centering
        \includegraphics[width=\textwidth]{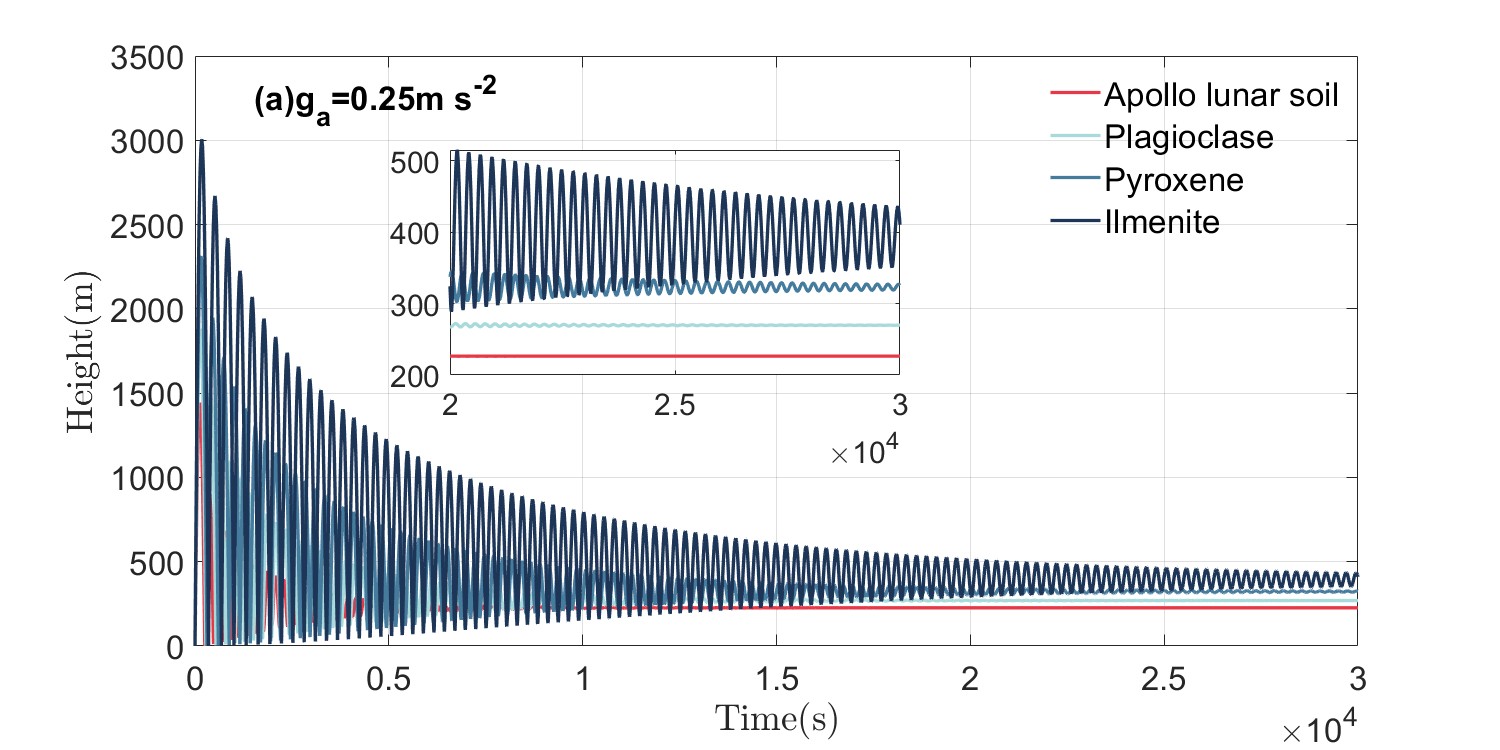}
    \end{minipage}
    \begin{minipage}{0.9\textwidth}
        \centering
        \includegraphics[width=\textwidth]{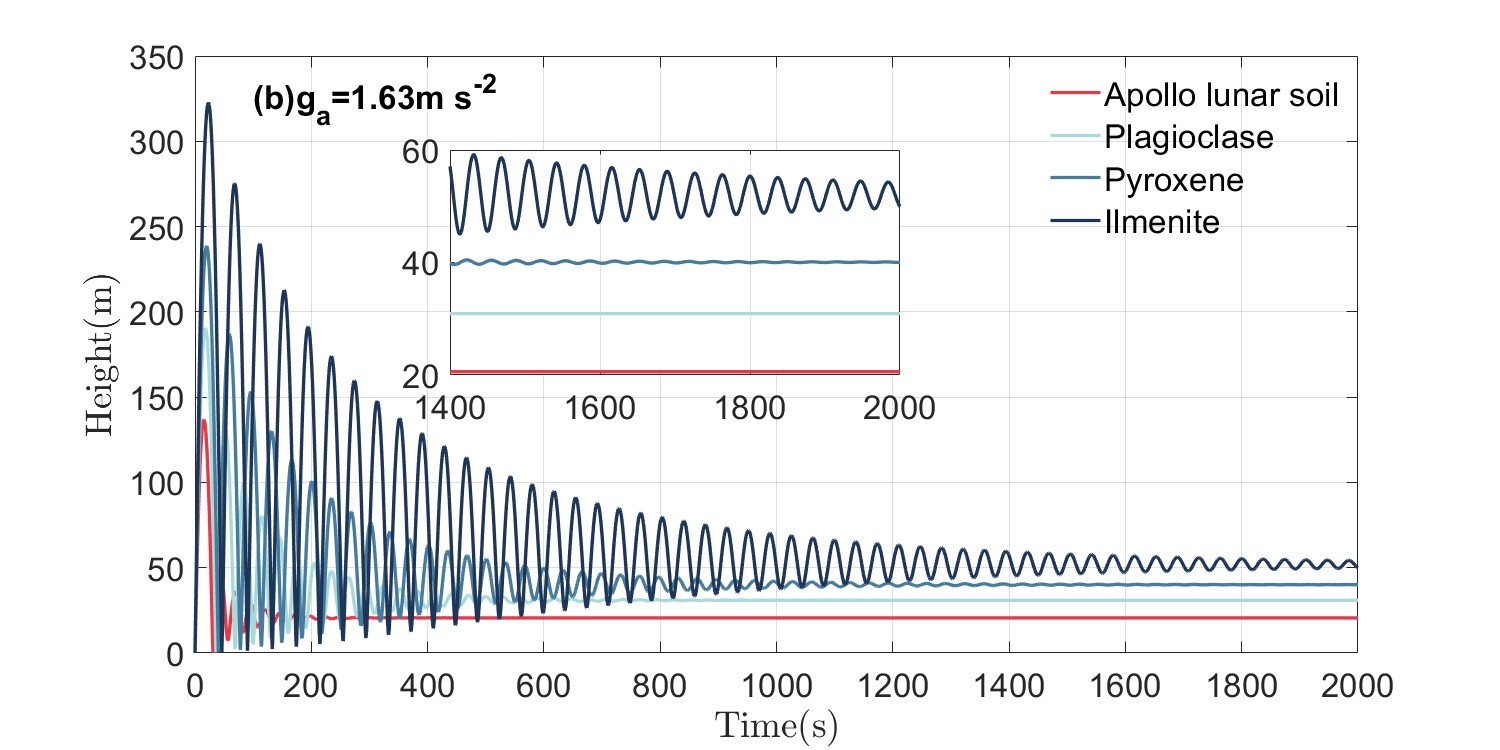}
    \end{minipage}
   \caption{(a) The vertical height of lofted dust particles as a function of time with $g_a=0.25m\;s^{-2}$, $\theta=0^{\circ}$, the results at 20000-30000 seconds have been magnified. (b) The vertical height of lofted dust particles as a function of time with $g_a=1.63m\;s^{-2}$, $\theta=0^{\circ}$, the results at 1400-2000 seconds have been magnified.}
   \label{fig.6}
\end{figure}

It is known that the dynamics and charging process are coupled, which means they can influence each other. Therefore, the effect of the work function on the charging process should naturally be inherited by the dynamics results. We solve Eqs.(\ref{eq.7})-(\ref{eq.8}) to obtain the vertical height of lofted dust particles, as depicted in Figure \ref{fig.6}.

In Figure \ref{fig.6}, the results reveal that the one-dimensional dynamics of dust particles exhibit damped oscillations as well, ultimately stabilizing at a particular height. However, the equilibrium height varies among different types of dust particles. In Figure \ref{fig.6}(a), for instance, the Apollo soil (red solid line) exhibits an equilibrium height of approximately 230m, whereas the Ilmenite with the smallest work function (deep blue solid line), can reach an estimated equilibrium height of 400m. Just as expected, the results suggest that particles with larger work functions exhibit lower equilibrium heights. This is because the photoelectron concentration and the equilibrium charge both are larger in the areas with lower work functions. According to Eq.(\ref{eq.9}), the higher equilibrium heights can be reached. Comparing the results between Figures \ref{fig.6}(a) and \ref{fig.6}(b), it is evident that larger gravity results in lower heights of dust particles. However, the variation in work functions yields consistent effects on the numerical results, indicating that changes in gravity do not bring about substantial impacts.

Finally, the equilibrium of lofted dust particles under other solar zenith angles(SZA) is shown in Figure \ref{eq.7}. In Figure \ref{eq.7}, with the same gravitational acceleration, the equilibrium height and charge numbers of the dust particles follow the trend that particles with larger work functions exhibit smaller equilibrium states when SZA varies from 0 to 90 degrees, while those with smaller ones exhibit larger equilibrium states. The equilibrium states of the same type of dust particle are larger with lower gravitational accelerations and smaller at higher ones, consistent with the earlier analysis. In addition,
although we only estimated the currents at noon in Figure (\ref{fig.3})-(\ref{fig.4}), owing to the equilibrium charge numbers correlating with the currents, and the magnitude of the equilibrium charge numbers hold an inverse relationship with the work functions when SZA varies from 0 to 90 degrees, we can extrapolate that the inverse relationship between the equilibrium currents and the work function also holds for the range of SZA from 0 to 90 degrees.

\begin{figure}[ht]
    \centering
    \begin{minipage}{0.90\textwidth}
        \centering
        \includegraphics[width=\textwidth]{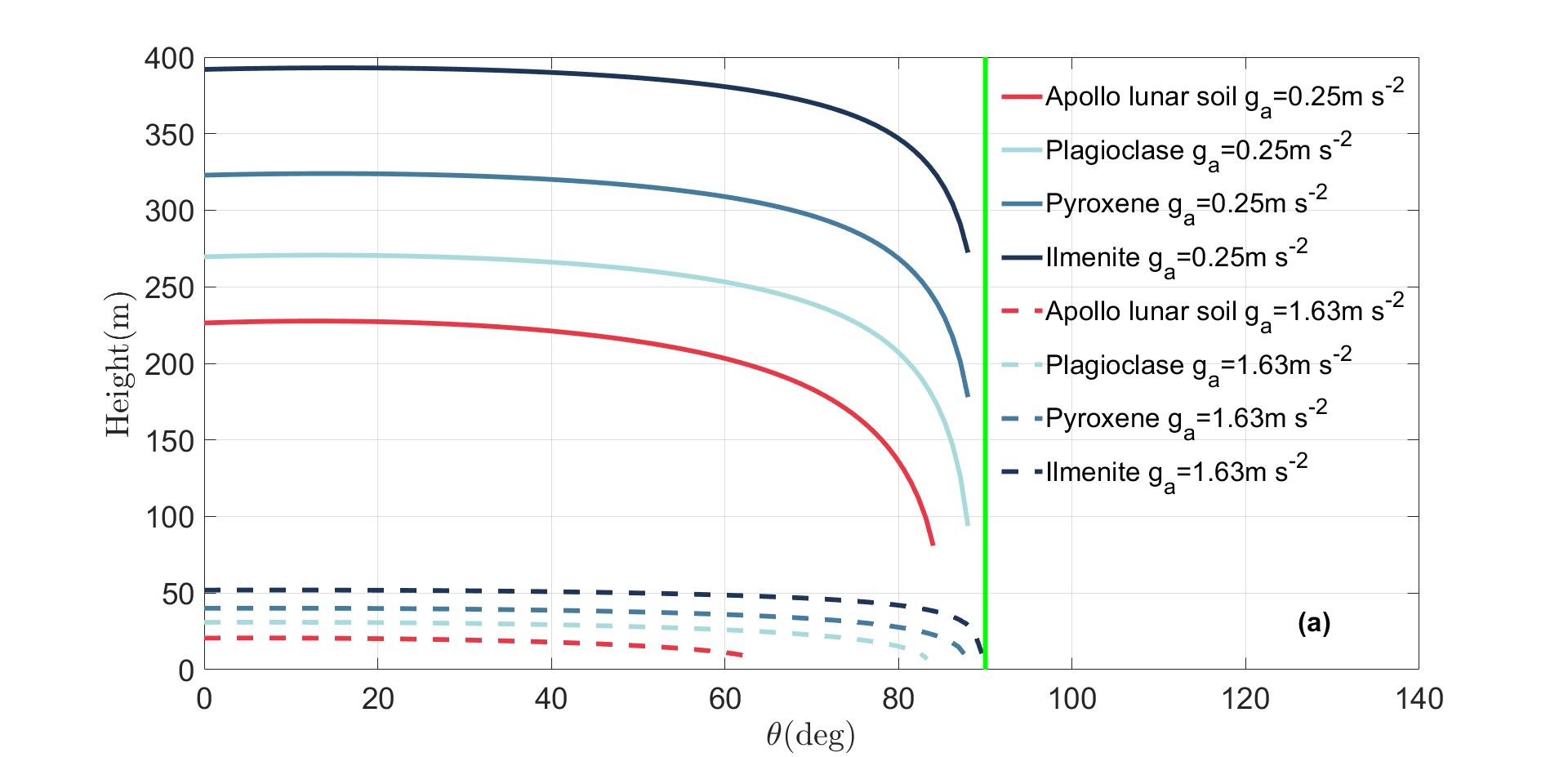}
    \end{minipage}
    \begin{minipage}{0.9\textwidth}
        \centering
        \includegraphics[width=\textwidth]{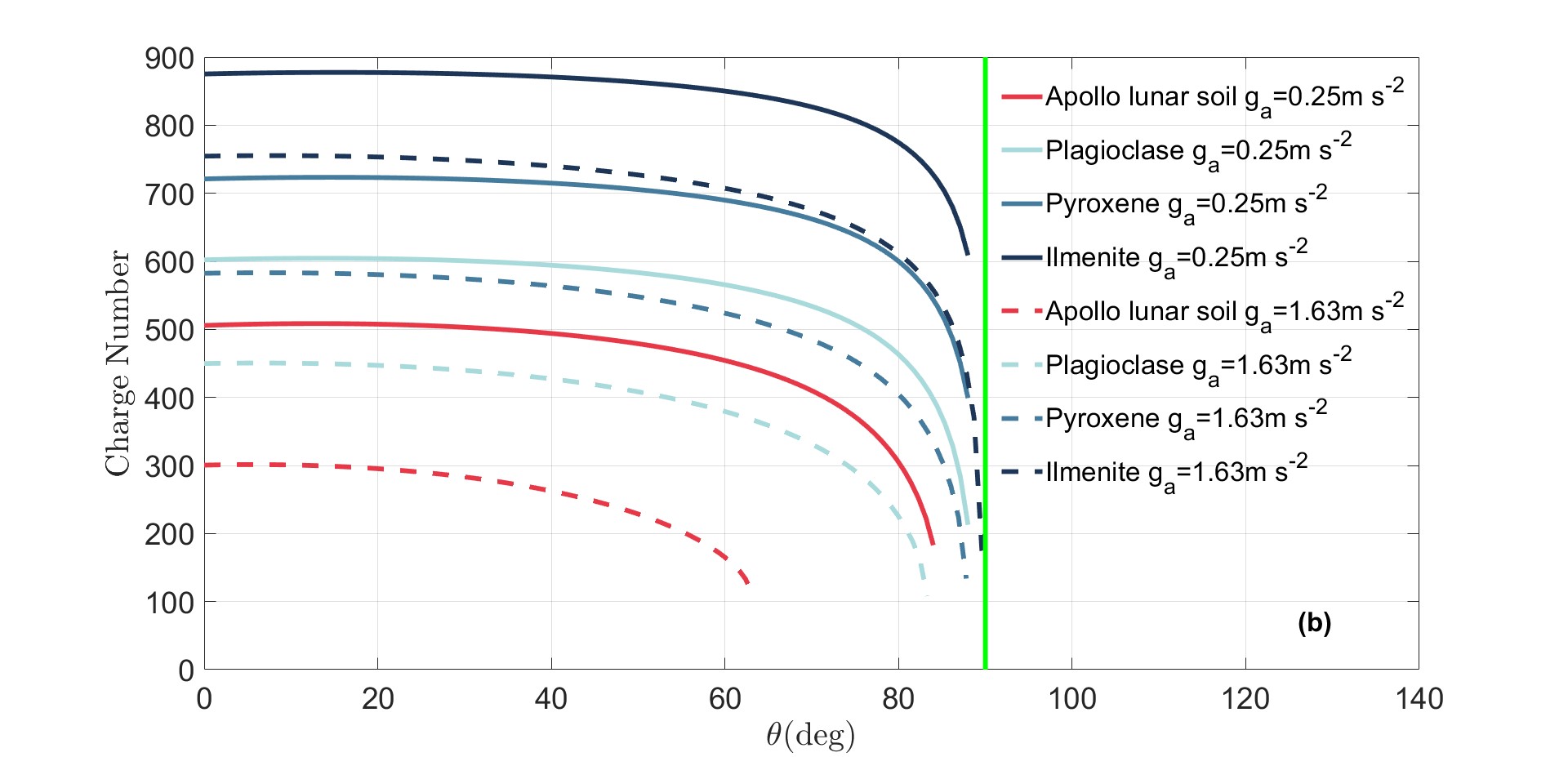}
    \end{minipage}
   \caption{(a) The equilibrium height of lofted dust particles with SZA varying from  $0^{\circ}$ to $90^{\circ}$. (b) The equilibrium charge numbers of lofted dust particles with SZA varing from $0^{\circ}$ to $90^{\circ}$. Both in (a) and (b), the solid line represents $g_a=0.25m\;s^{-2}$, and the dash line represents $g_a=1.63m\;s^{-2}$, and the green solid line represents $\theta=90^{\circ}$.}
   \label{fig.7}
\end{figure}

Additionally, it is observed that the equilibrium states of the dust particles exhibit a cutoff SZA where they rapidly decrease. Taking the Apollo lunar soil as an example, the cutoff SZA is around 60 degrees when $g_a=0.25m\;s^{-2}$, and around 80 degrees when $g_a=1.63m\;s^{-2}$. This suggests that dust particles fail to reach equilibrium and eventually return to the surface at the cutoff SZA. This can be attributed to the fact that when SZA increases, a decrease in the concentration of photoelectrons can result in the electrostatic force to be insufficient to support the stable levitation. It is worth noting that the cutoff SZA of dust particles with a lower work function is larger than that of those with a higher work function, which follows the same relationship with work function like the equilibrium states do. This inverse relationship can be seen more clearly in Figure \ref{fig.7}(a). As mentioned above, a decrease in the concentration of photoelectrons can result in the electric field force to be insufficient to balance the gravity. Due to the poor ability to emit photoelectrons of the areas with a high work function, the concentration of photoelectrons in these areas will reach the threshold more quickly that cannot support stable levitation when the SZA increases. This leads to the earlier decrease in the equilibrium heights(Fig.\ref{fig.7}(a)) and equilibrium charge numbers(Fig.\ref{fig.7}(b)).

\section{conclusion} \label{sec:conclusion}
we study the charging and dynamics of four types of dust particles with different work functions and two different gravitational acceleration values. The numerical results reveal that the significant difference in the concentration of photoelectrons of the dust charging areas due to the change in work function will result in different equilibrium currents, equilibrium heights, equilibrium charges, and cutoff SZA at which the dust particles cannot stably levitate. Our results support the conclusion that even if the gravitational acceleration changes, the equilibrium states all hold a clear inverse relationship with the work functions of dust particles when the SZA varies from 0 to 90 degrees except the equilibrium solar ion currents. The cutoff SZA also holds an inverse relationship with the work function.

In this paper, we do not consider the effects of celestial rotation and topography on the plasma environment, and future studies can investigate the charging mechanism of dust particles on the surface of irregularly rotating airless celestial bodies. However, our conclusion implies that layered dust clouds with different electrostatic environments may exist within a few hundred meters from the surface of an airless celestial body\citep{2013ApJ...771L..36J,2021PSJ.....2...67B}, which means that landing spacecraft are subject to different influences from charged dust at different altitudes, including mechanical abrasion, electrostatic interference, etc. Our study might help with the design of spacecraft for dust and antistatic protection and a better understanding of the dust environment on the surface of airless celestial bodies.


\begin{acknowledgments}
This work was supported by the National Natural Science Foundation of China (grant No. 42241148 and No. 51877111).
\end{acknowledgments}

\bibliography{sample631}{}
\bibliographystyle{aasjournal}

\end{document}